\documentclass[amsmath,10pt,twocolumn,superscriptaddress,footinbib,pre]{revtex4-1}
\usepackage{amsmath,amsfonts,bm,dsfont}
\usepackage{graphicx}
\usepackage{subfigure}
\usepackage{float}
\usepackage{siunitx}
\usepackage{uri}

\usepackage[dvipsnames,usenames]{xcolor} 

\usepackage{color}
\usepackage[colorlinks=true,allcolors=blue]{hyperref}
\usepackage{tikz}

\begin{document}
\title{Using tensor network states for multi-particle Brownian ratchets}
\author{Nils E.~Strand}
\author{Hadrien Vroylandt}
\author{Todd R.~Gingrich}
\email{todd.gingrich@northwestern.edu}
\affiliation{Department of Chemistry, Northwestern University, 2145 Sheridan Road, Evanston, Illinois 60208, USA}

\begin{abstract}
  The study of Brownian ratchets has taught how time-periodic driving supports a time-periodic steady state that generates nonequilibrium transport.
  When a single particle is transported in one dimension, it is possible to rationalize the current in terms of the potential, but experimental efforts have ventured beyond that single-body case to systems with many interacting carriers.
  Working with a lattice model of volume-excluding particles in one dimension, we analyze the impact of interactions on a flashing ratchet's current.
  To surmount the many-body problem, we employ the time-dependent variational principle with a binary tree tensor network, methods discussed at length in a companion paper.
  Rather than propagating individual trajectories, the tensor network approach propagates a distribution over many-body configurations via a controllable variational approximation.
  The calculations, which reproduce Gillespie trajectory sampling, identify and explain a shift in the frequency of maximum current to higher driving frequency as the lattice occupancy increases.
\end{abstract}
\maketitle

\textit{Introduction}.---The rectification of thermal noise into directed motion via a ratchet effect has attracted sustained fascination and attention~\cite{smoluchowski1927experimentell,feynman2011feynman,harmer2001brownian,reimann2002brownian}.
Theoretical studies of that ratcheting effect are dominated by models of one-body dynamics driven by a time-dependent potential, with a particular focus on how the generated current varies with system parameters like a driving frequency~\cite{bartussek1994periodically,elston1996numerical,reimann2001introduction} or the diffusion constant~\cite{buttiker1987transport,kostur2001multiple,tammelo2002three,zeng2010current}.
These models have captured the essential ratcheting effect and have guided experimental efforts, but they leave out an important control that experimentalists have over their systems: the collective effects that can emerge out of interactions between multiple ratcheted particles~\cite{kedem2019cooperative,kodaimati2019empirical,lau2020electron,Craig2006,Li2016a}.
Similar collective effects are known to generate rich dynamical behavior in nonequilibrium steady states (NESS), for example in coupled molecular motors~\cite{Klumpp2005, Imparato2015, Golubeva2012, Campas2006, Vroylandt2020b} and in the asymmetric exclusion process (ASEP)~\cite{chou2011non,lazarescu2015physicist}.
In the case of the ASEP, powerful numerical methods built upon tensor networks have been employed to interrogate those dynamic phase transitions~\cite{helms2019dynamical, helms2020dynamical,Proeme2010}, opening the door to also consider the impact of interactions on time-periodic steady states in many-particle ratchets.

In this letter and in a companion paper, we develop and apply those tensor network tools to compute steady-state currents in a multi-particle 1D ratchet.
Our approach mixes the spectral large-deviation theoretic analysis of one-body ratchets we previously reported~\cite{strand2020current} with the quantum dynamics literature's tensor network methods~\cite{schollwock2011density,paeckel2019time} which are presently finding applications to classical stochastic dynamics~\cite{helms2019dynamical,banuls2019using,helms2020dynamical,Nagy2002,Hieida1998,Temme2010,Johnson2010,Johnson2015}.
Those tensor network tools are essential for handling the many-body problem because more traditional matrix algebra techniques cannot be applied when the state space grows exponentially with the number of interacting particles.
In the quantum many-body context, the density matrix renormalization group (DMRG) technique~\cite{white1992density} solves for ground states of spin lattices while the time dependent variational principle (TDVP) technique propagates a many-body state in time~\cite{haegeman2011time,haegeman2016unifying,bauernfeind2020time,Yang2020a,paeckel2019time,kloss2018time}.
For classical NESS stochastic dynamics, a scaled cumulant-generating function (SCGF) is analogous to the quantum ground state energy calculation, so DMRG has been applied to calculate the SCGF for currents in the 1D and 2D ASEP~\cite{helms2019dynamical,helms2020dynamical,Gorissen2012,Gorissen2011,Lazarescu2013} as well as in other kinetically constrained models~\cite{banuls2019using,Causer2021optimal,Causer2021finite}.

Unlike those NESS problems, here we consider a time-dependent steady state generated by a time-dependent rate matrix. 
As such, it is not sufficient to use DMRG to compute the SCGF as an eigenvalue optimization problem.
Instead, we use TDVP to propagate a tensor network state until a time-periodic steady state is reached and we extract the SCGF for the period-averaged current from this long-time limit.
We focus this work on a flashing ratchet with a square wave driving, but we highlight that the application of TDVP to evolve many-body classical stochastic dynamics is straightforwardly extended to arbitrary driving potentials.

\begin{figure}[htb]
  \centering
  \includegraphics[width=0.5\textwidth]{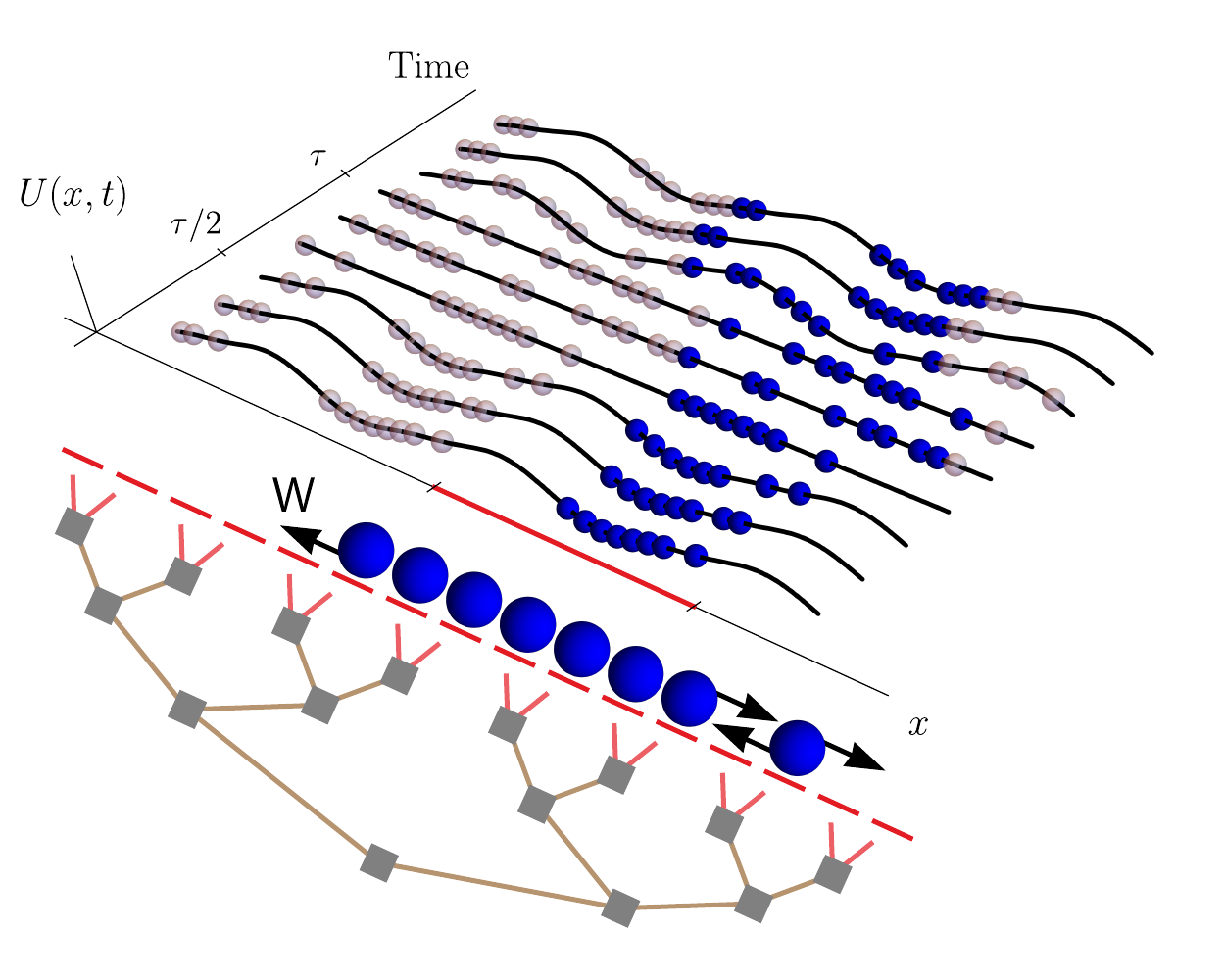}
  \caption{\emph{Top.} Volume-excluding particles (blue) and their periodic replicas (light purple) in a one-dimensional flashing ratchet.
    The impact of interactions on the current can be ascertained by following individual trajectories.
    \emph{Bottom.} Alternatively, a superposition of one or more many-body configurations can be encoded in a binary tree tensor network, formed from gray tensors \(\left[A\right]\), brown auxiliary indices with maximum bond dimension \(m\), and red physical indices that take the values one or zero to reflect occupied or unoccupied sites.
    For sufficiently large \(m\), the tree tensor network's time evolution approximates the propagation of the initial probability density under a time-dependent rate matrix \(\mathsf{W}\).}
  \label{setup}
\end{figure}
\textit{Discrete-state ratchet model}.---The tensor network methods thrive on discrete-state models.
While we imagine a 1D flashing ratchet as being generated from turning on and off a smooth continuous potential as in Fig.~\ref{setup}, we immediately discretize that potential onto an \(N\)-site lattice with periodic boundary conditions, following our prior work~\cite{gingrich2017inferring,strand2020current}.
Dynamics of that discrete-space model is governed by the master equation, \(\partial|p\rangle/\partial t=\mathsf W |p\rangle\).
Here, \(|p\rangle\) is a state vector consisting of configurational probabilities and \(\mathsf W\) is a time-varying rate matrix, with rates constructed so the continuous-space limit alternates between evolution on a potential \(U(x)= - V_{\rm max} \left[\frac{a_1}{2} \sin\left(\frac{2 \pi x}{x_{\rm max}}\right) + \frac{a_2}{2} \sin\left(\frac{4 \pi x}{x_{\rm max}}\right)\right]\) for time \(\tau / 2\) and evolution on a flat potential for time \(\tau / 2\).
The tunable parameters \(V_{\rm max}\), \(a_1,\) and \(a_2\) sculpt the form of the biharmonic potential while \(x_{\rm max}\) sets a box length and \(\tau\) a period of driving.
Corresponding to the scales of some experimental Brownian ratchets~\cite{kedem2017light}, we set \(x_{\rm max} = 1~\si{\um}, a_1 = 1, a_2 = 0.25,\) and \(V_{\rm max} = 0.1~\si{\V}\).

In the rate matrix description, the dynamics involves toggling between two constant-in-time rate matrices, \(\mathsf W_1\) and \(\mathsf W_2\).
The flat-potential rate matrix \(\mathsf W_2\) has rates between neighboring lattice sites given by \(r_{2,i\rightarrow i\pm 1}= D/h^2\), in terms of the lattice spacing \(h = x_{\rm max} / N\) and a diffusion constant \(D\), taken to be \(12.64~\si{\um\squared\per\ms}\) except where otherwise noted.
The other rate matrix has nearest-neighbor transitions with the same diffusive contribution but with a drift term as well: \(r_{1, i\rightarrow i\pm 1}=\pm U'(x) / 2h + D/h^2\).
As in the ASEP, we construct an exclusion process, so while the lattice can have more than one particle, each lattice site can house at most one particle.
Consequently, the rate \(r_{k,i \to i\pm 1}\) is zero if site \(i\pm 1\) is already occupied.
As always, diagonal elements of the rate matrices are set to ensure that columns sum to zero and probability is conserved~\cite{vankampen2007stochastic}.

Our focus is on the impact of the exclusion interactions on the period-averaged current \(\bar{\jmath}=(1/\tau)\int_0^\tau\text{d}t\,\sum_{i} j_{i\to i+1}(t)\), where \(j_{i \to i+1}(t)\) is the current from site \(i\) to site \(i+1\) at time \(t\).
Due to the periodic boundary conditions, we associate \(N+1 \equiv 1\).
The statistics of \(\bar{\jmath}\) can be extracted from the SCGF, \(\psi_{\bar{\jmath}}(\lambda)=\lim_{n\to\infty}(\ln\langle e^{\lambda n\bar{\jmath}}\rangle_n/n)\), where \(\langle \cdot \rangle_n\) denotes an average over trajectories with \(n\) driving periods.
For example, the mean is the first derivative at \(\lambda = 0\): \(\left<\bar{\jmath}\right> = \psi'(0)\).
We compute \(\psi(\lambda)\) by first constructing tilted rate matrices \(\mathsf W_k(\lambda)\) with modified jump rates \(r_{k, i \to i\pm 1}(\lambda) = r_{k, i\to i \pm 1} e^{\pm \lambda}\).
The tilted rate matrices' diagonal elements are unmodified from \(\mathsf W_k\), so \(\mathsf W_k(\lambda)\) are not themselves rate matrices.
Rather, they generates dynamics with an extra exponential bias on the current~\cite{lebowitz1999,Lecomte2007}.
For the NESS, it has become standard practice to compute \(\psi(\lambda)\) from the maximal eigenvector of \(\mathsf{W}(\lambda)\).
Because the NESS tilted rate matrix shares eigenvectors with the tilted propagator \(e^{\mathsf{W}(\lambda) t}\), the SCGF can be computed without evolving dynamics.
For the time-periodic steady state, we instead compute \(\psi(\lambda)\) as the maximal eigenvalue of the full-period tilted propagator \(\mathsf{T} \equiv e^{\mathsf{W_2}(\lambda) \tau / 2} e^{\mathsf{W_1}(\lambda) \tau / 2}\), extracted using the power iteration method~\cite{strand2020current}.
An arbitrary initial state vector \(\left|p\right>\) is evolved via TDVP for many periods to reach a normalized steady state \(\left|\pi(\lambda)\right>\), at which point the SCGF measures the rescaling of \(\left|\pi(\lambda)\right>\) under one more period of dynamics:
\begin{equation}
  \psi(\lambda) = \ln \left<\pi(\lambda)|\mathsf{T}(\lambda)|\pi(\lambda)\right> / \tau.
  \label{scgf}
\end{equation}

\textit{Tensor network structure and methods}.---For single-body dynamics, \(\left|p\right>\) may be sufficiently low dimensional that the \(\mathsf{W}_k(\lambda)\) are constructed as explicit matrices and the time-propagation is computed with a numerically evaluated matrix exponential~\cite{strand2020current}.
 As more particles are added to the ratchet model, that direct calculation becomes intractable, demanding the time evolution be executed with a tensor network approximation.
 Before making any approximations, a vector in the full state space can be expanded in terms of local basis states as \(\left|p\right> = \sum_{s_1, \hdots,s_N} c_{s_1, \hdots,s_N} \left|s_1\hdots s_N\right>\),
 where the coefficient tensor \(c_{s_1\cdots s_N}\) expresses how the state is built up as a superposition of states \(\left|s_1 \hdots s_N\right>\) with \(s_1\) particles in site 1, \(s_2\) particles in site 2, \(\hdots\), and \(s_N\) particles in site \(N\).
 The tensor network approximation restricts the correlations between sites by requiring that the coefficient tensor be constructed as a partially contracted binary tree formed from a set of smaller-rank tensors \(\left[A\right]\) (see Fig.~\ref{setup}).
 With \(N\) dangling legs (red), that network is a rank-\(N\) tensor whose physical indices indicate whether a lattice site is occupied.
 The tensor network additionally consists of many contracted auxiliary indices (brown) shared by two tensors, and the largest dimension \(m\) of those auxiliary indices serves as a variational parameter which controls the extent to which information is communicated from tensor to tensor.
 A choice of \(m\) defines a variational subspace spanned by states that are parameterized by the set of tensors used to construct the coefficient tensor: \(\left|p[A]\right>\).

    \begin{figure*}[htb]
    \centering
    \begin{tikzpicture}
        \node at (0,0) {\includegraphics[width=0.45\textwidth]{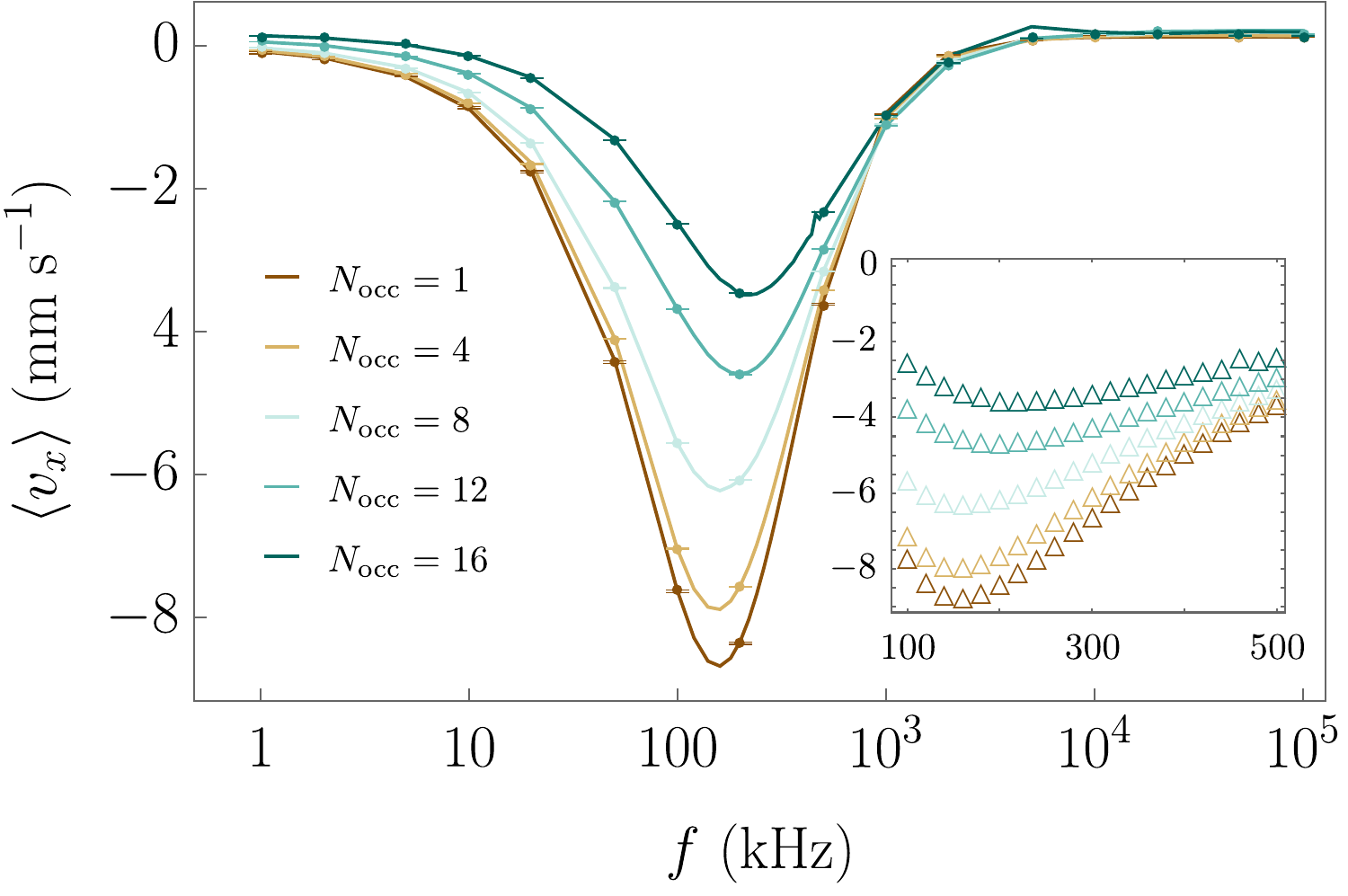}};
        \node at (9,0.1) {\includegraphics[width=0.46\textwidth]{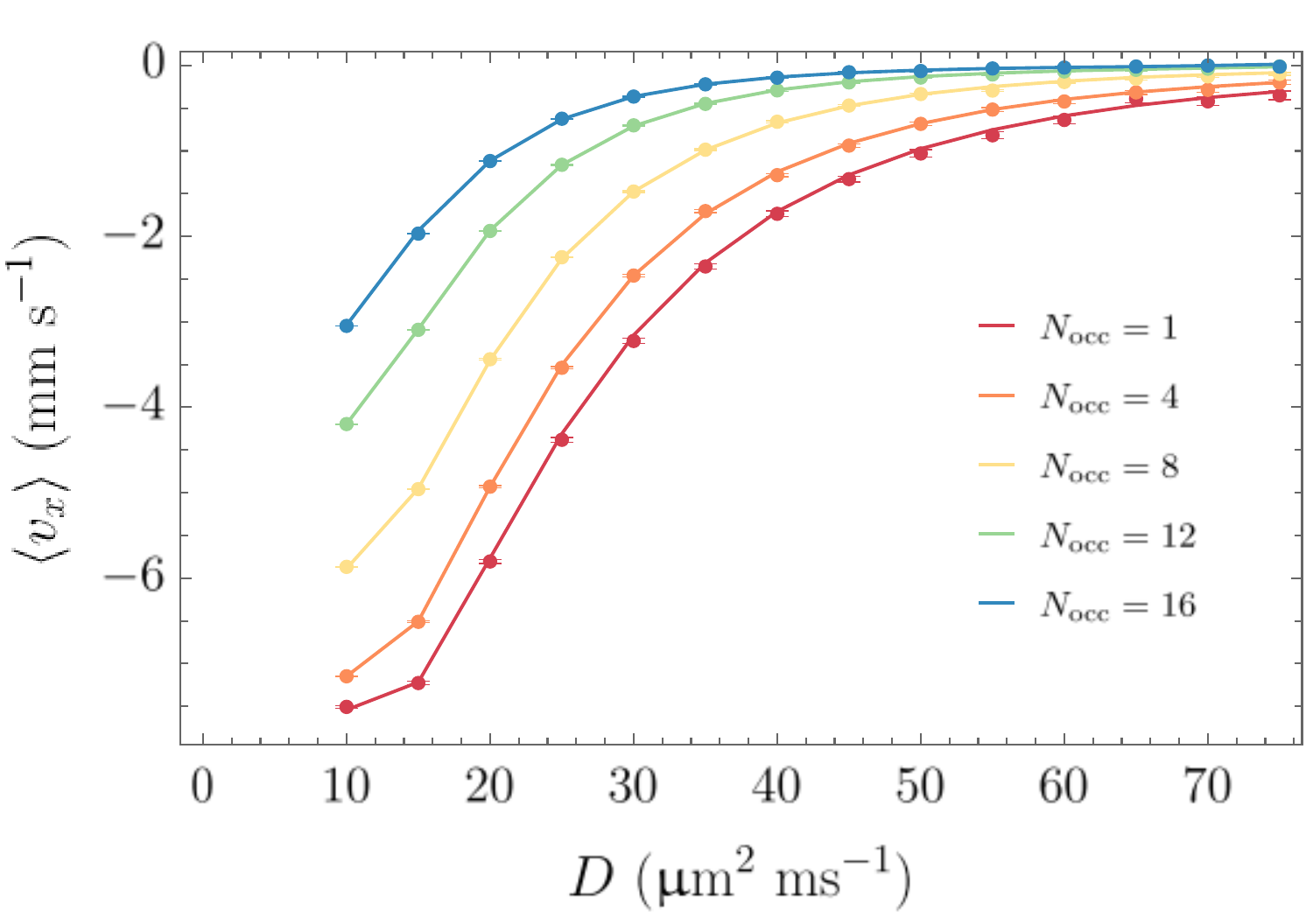}};
        \node at (-4,2.5) {(a)};
        \node at (5,2.5) {(b)};
    \end{tikzpicture}
    \caption{Mean per-particle current induced by ratcheting \(N_{\rm occ}\) volume-excluding particles on a 32-site lattice.
      Currents were computed by the TDVP SCGF method using timestep \(\Delta t = 1~\si{\ns}\), \(\delta = 10^{-4}\), and maximum bond dimension \(m = 200\) (lines) and were found to agree with Gillespie sampling (points, one standard error of the mean error bars).
      \emph{Left.} Current vanishes in the low- and high-frequency limit.
      Current is induced at intermediate frequencies, with the frequency of maximal current shifting higher as particles are added.
      The inset shows TDVP data that highlight that shift.
      \emph{Right.} Agreement between TDVP and Gillespie persisted across a range of diffusion coefficients, plotted here with \(f = 100~\si{\kHz}\).}      
    \label{jresults}
 \end{figure*}
 TDVP evolves the tensor network state \(\left|p[A]\right>\) with the restriction that each step of the evolution remains in the variational subspace so that at every moment in time, the new state can also be written in the form \(\left|p[A']\right>\) for some new set of tensors \(A'\).
 Though more commonly utilized with a matrix product state (MPS), TDVP has been previously developed for the binary tree tensor network (BTTN)~\cite{kohn2020superfluid,bauernfeind2020time,kloss2020studying} that we employ.
 The algorithm, discussed in significantly greater detail in a companion paper~\cite{strand2021companion}, consists of sweeps that pass through the tree to advance each tensor of the set \([A]\) by one timestep \(\Delta t\).
 It is crucial the the tree's rank-3 tensors can be efficiently updated one at a time, an efficiency gained because a gauge freedom can be leveraged to bring the loopless BTTN into a canonical form.
 By contrast, an MPS treatment would require a looped MPS, and the loops in the tensor network are known to degrade computational performance~\cite{schollwock2011density,pippan2010efficient}.
 It is also important that the action of \(\mathsf{W}_k(\lambda)\) on \(\left|p[A]\right>\) can be computed without ever constructing the explicit tilted rate matrices.
 In lieu of constructing the matrix form of the operators, they are cast in a second quantized form as
 \begin{align}
\nonumber    \mathsf W_k(\lambda)=&\sum_{i=1}^N r_{k,i\rightarrow i+1}(e^\lambda \mathbf a_i\mathbf a_{i+1}^\dagger-\mathbf n_i\mathbf v_{i+1})\\
    +&\sum_{i=1}^Nr_{k,i+1\rightarrow i}(e^{-\lambda}\mathbf a_i^\dagger \mathbf a_{i+1}-\mathbf v_i\mathbf n_{i+1}),
    \label{eq:secondquant}
 \end{align}
 where \(\mathbf a_i\), \(\mathbf a_i^\dagger\), \(\mathbf n_i\), and \(\mathbf v_i\) are fermionic annihilation, creation, particle number, and vacancy number operators at site \(i\), respectively.
 The nearest-neighbor structure of \(\mathsf{W}_k(\lambda)\) allows the operators to be factorized into a matrix product operator (MPO) that associates to each physical index a low-rank tensor.
 Action of \(\mathsf W_k(\lambda)\) on a state \(\left|p[A]\right>\) is practically computed by contracting over the physical indices that connect the MPO to the BTTN state.

 \textit{Calculations}.--- Calculations with \(N_{\rm occ}\) particles spread over \(N\) lattice sites were initialized in a random pure state that placed the particles into sites in an arbitrary manner.
 Starting from that initial seed, the (\(\lambda = 0\)) steady-state of \(\mathsf{W_2}\), \(\left|\pi_2\right>\), was reached by DMRG.
 Though the initial seed has a trivial bond dimension of one, a version of DMRG that performs subspace expansion~\cite{hubig2015strictly} was employed to systematically grow the bond dimension such that \(\left|\pi_2\right>\) would have a maximum bond dimension of \(m\).
 As \(\mathsf{W_2}\) is itself a rate matrix, the eigenvalue associated to \(\left|\pi_2\right>\) is zero, making clear when the DMRG procedure was fully converged.
 Starting from this \(\left|\pi_2\right>\), the TDVP algorithm evolved the BTTN state in time subject to tilted propagators \(\mathsf{W_1}(\lambda)\) and \(\mathsf{W_2}(\lambda)\) for very small positive and negative biasing values \(\lambda = \pm\delta\).
 Sufficiently many periods of driving were evolved so the BTTN state converged to \(\left|\pi(\pm \delta)\right>\), and the SCGF could be computed from Eq.~\eqref{scgf}.
 The mean current was then computed as the finite difference \(\left<\bar{\jmath}\right> \approx (\psi(\delta) - \psi(-\delta)) / 2\delta\).
 Our companion paper discusses the convergence properties of both DMRG and TDVP algorithms, which were implemented using the ITensor library~\cite{itensor}.
 There we demonstrate that, provided the bond dimension \(m\) is sufficiently large, the mean current agrees with that of trajectories sampled from a version of the Gillespie algorithm for time-dependent driving~\cite{strand2021companion,anderson2007modified}.
 For an \(N = 32\) lattice, \(m = 200\) is sufficient, allowing us to analyze how the many-particle ratchets respond to both the driving frequency \(f \equiv 1 / \tau\) and the diffusion constant. 
 
 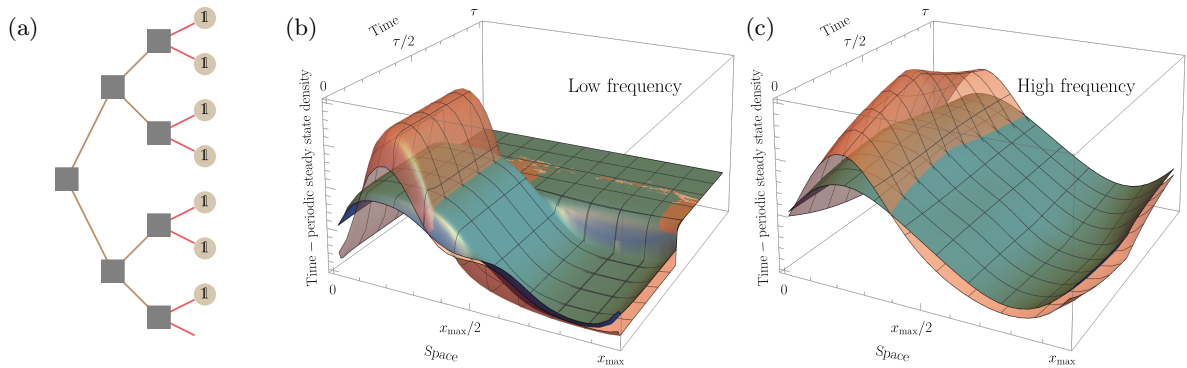
\begin{figure*}[htb]
   \centering
   \begin{tikzpicture}
     \node at (0,0) {\includegraphics[width=0.12\textwidth]{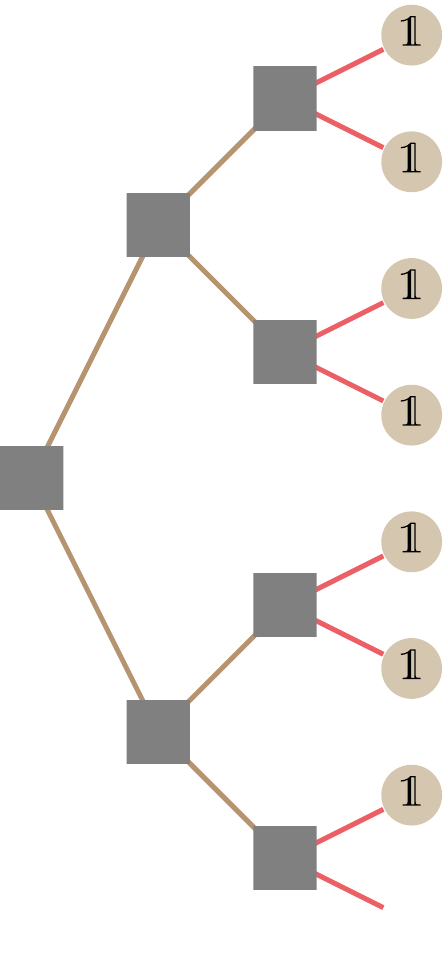}};
     \node at (5,0) {\includegraphics[width=0.35\textwidth]{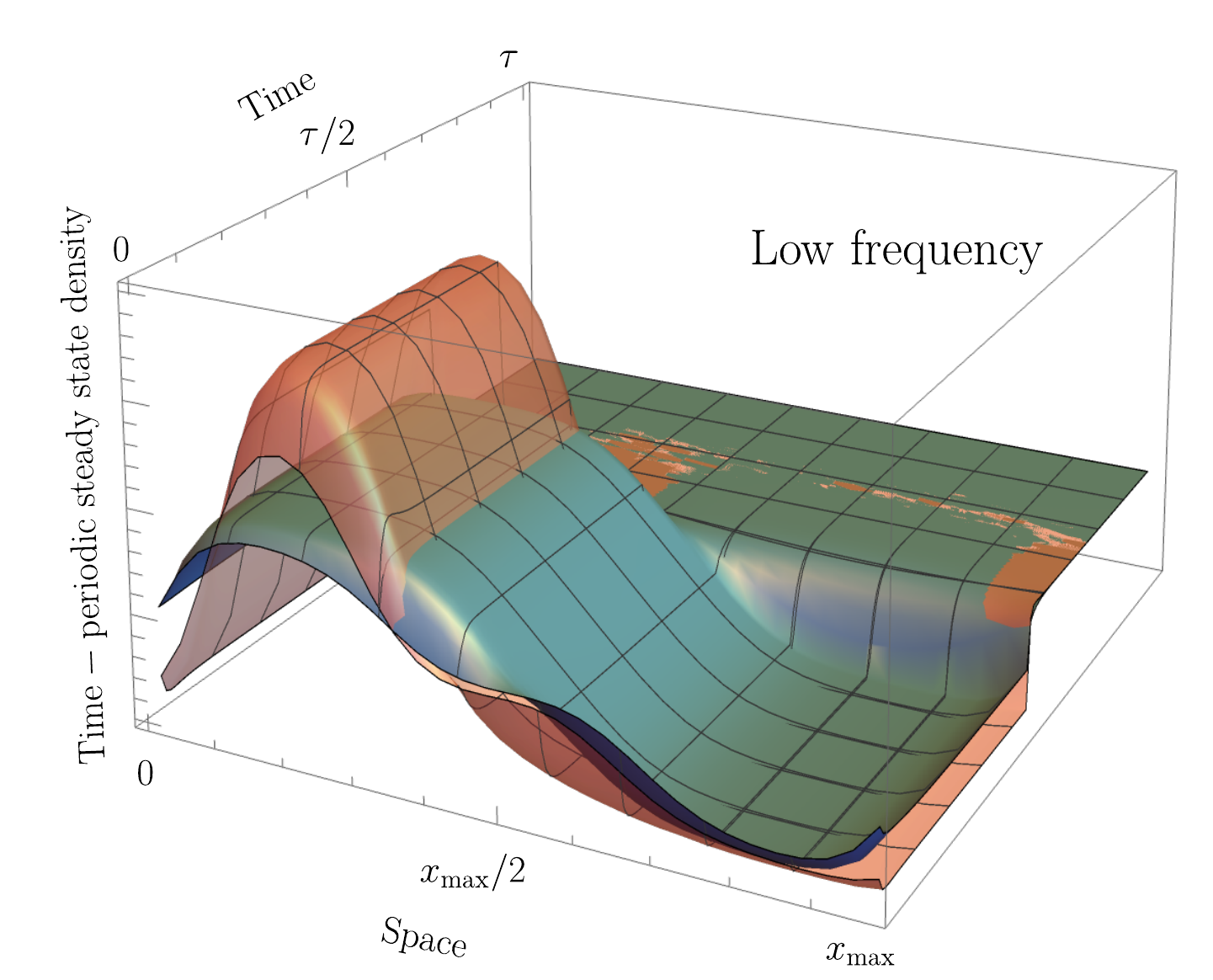}};
     \node at (11,0) {\includegraphics[width=0.35\textwidth]{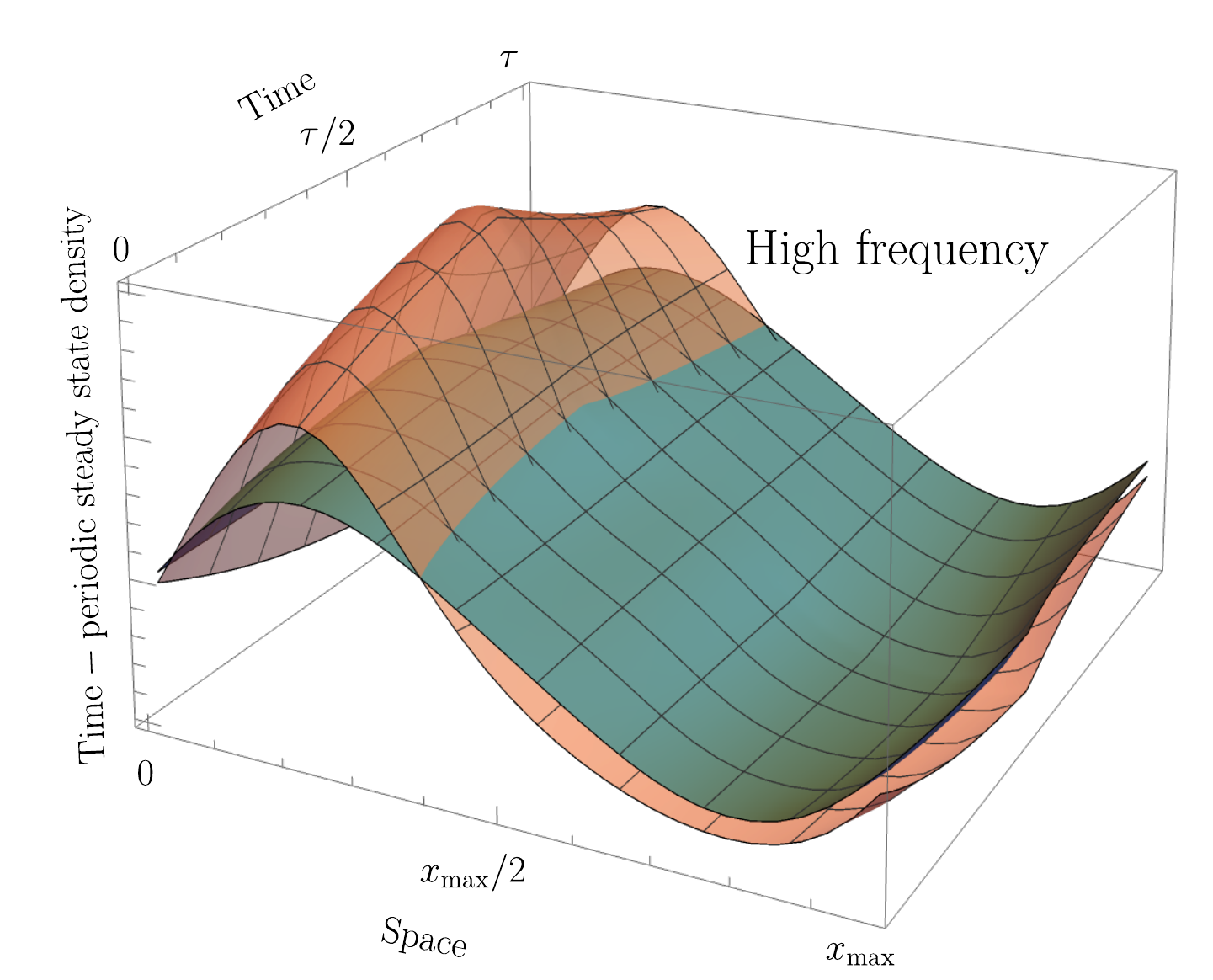}};
     \node at (-1.5,2) {(a)};
     \node at (2.2,2) {(b)};
     \node at (8.3,2) {(c)};
     
   \end{tikzpicture}
   \caption{\emph{Left.} By contracting all but site \(i\) with a vector of ones, the time-periodic steady state BTTN can be efficiently traced over to give the steady-state probability of finding a particle at site \(i\).
     \emph{Center.} Performing that trace for all sites and at all times gives the spatio-temporal evolution of particle density on a 32-site lattice.
     At low frequency (10~\si{kHz}), the density is approximately the equilibrium Boltzmann distribution with particles filling the potential well in the first half-period of driving and distributing themselves uniformly in the second half-period.
     Due to particle interactions, the \(N_{\rm occ} = 16\) (green) distribution must spread the particles across the entire well whereas they congregate more narrowly in the bottom of the well for \(N_{\rm occ} = 8\) (orange).
     \emph{Right.} At high frequency (1~\si{MHz}), the temporal variation in the density is smoothed out.
     Increasing \(N_{\rm occ}\) again forces density to spread more broadly, but it is not straightforward to relate densities to energy wells when the particles move on a comparable timescale as the driving frequency.}
   \label{densresults}
 \end{figure*}

 \textit{Results}.---We illustrate those responses by computing a particle's period-averaged velocity \(\left<v_x\right>\) as a function of \(N_{\rm occ}\).
 Since each particle is statistically equivalent, \(\left<v_x\right>=\left<\bar{\jmath}\right>h/N_{\rm occ}\), and \(\left<v_x\right>\) can be interpreted as a period-averaged current per particle.
 Plots of that per-particle current in Fig.~\ref{jresults} show excellent agreement with the Gillespie sampling, capturing even the effects of interactions as the carrier density is increased.
 
  The single-particle \(N_{\rm occ} = 1\) case recovers the well-studied flashing ratchet behavior.
  At both low- and high-frequency driving, the current vanishes, but leftward motion is induced at intermediate frequencies due to the asymmetric shape of the potential.
  Each potential well is bordered by a close barrier to the left and a far barrier to the right, barriers that go away during the free diffusion stage of the driving protocol.
  The intermediate driving frequency allows the particle's \(\mathsf{W_2}\) free diffusion to make it past the left barrier more frequently than it passes the right, hence the current.
  At a fixed frequency, that current is amplified by tuning \(D\) such that the typical diffusion length exceeds the distance from the well to the left barrier but not to the right. 
  Our work supplements this single-particle understanding to include the interacting neighbors whose presence frustrate the flow.

  That frustration decreases each particle's effective diffusion constant, but in a manner which is qualitatively distinct from merely tuning \(D\) to smaller values.
  To appreciate the difference, imagine starting with the \(N_{\rm occ} = 1\) ratchet with \(D = 40~\si{\um\squared\per\ms}\) and slowing the diffusion either by decreasing \(D\) directly or by increasing \(N_{\rm occ}\).
  Whereas decreasing \(D\) can enhance the asymmetry and amplify per-particle current, slowing diffusion by jamming always attenuates the per-particle current.
  Fig.~\ref{jresults}b shows that attenuation across a range of diffusion constants.

  The jamming-induced drop in per-particle current persists across different driving frequencies, but Fig.~\ref{jresults}a shows the extent of the current attenuation is frequency dependent.
  Consequently, the frequency of maximal current shifts higher as \(N_{\rm occ}\) increases.
  To understand that shift, it is useful to convert the time-evolved BTTN state into a spatio-temporal evolution of particle density.
  Though our initial motivation for the BTTN TDVP was to compute the mean scalar current \(\left<\bar{\jmath}\right>\), the converged BTTN contains much more information about the dynamics.
  For example, all but one site of the BTTN can be traced over to leave the probability of occupying the remaining site.
  Fig.~\ref{densresults} shows the particle densities computed by repeating the partial trace for all times and at all lattice sites.
  Those many-particle densities rationalize the frequency shift of Fig.~\ref{jresults}a.
  At low frequencies, the particles have time to relax into the potential well, concentrating them together so volume exclusion effects are significant.
  At higher frequencies, the density is more uniformly distributed throughout the period because the particles don't have enough time to settle in the bottom of the well before the potential is switched.
  Due to that more uniform density, the high-frequency ratchet is less strongly influenced by interparticle interactions, manifesting as a shift in the peak frequency.
  
  \textit{Discussion}.---We have demonstrated how a many-body 1D flashing ratchet responds to volume exclusion.
  In doing so, have highlighted that tensor networks offer a tantalizing new way to study classical, many-body time-dependent steady states.
  The time-dependent master equation we solve has traditionally been approached by sampling realizations of the process via the Gillespie algorithm or by propagating an initial distribution through time.
  The Gillespie approach is extremely flexible, though it becomes markedly more cumbersome for a time-dependent rate matrix~\cite{anderson2007modified}.
  As it is a Monte Carlo approach, it requires sampling a large ensemble of trajectories, and that ensemble must be resampled (or reweighted) if system parameters are altered.
  Evolving the distribution is appealing because it obviates the noise of trajectory sampling and effectively integrates over all trajectories\textemdash rare and typical, but the approach has always been limited by the size of the state space.
  Whereas the one-body calculation is readily solved via a matrix exponential, the ratchet rate matrix grows combinatorially with dimension \(N\) choose \(N_{\rm occ}\).
  For the \(N = 32, N_{\rm occ} = 16\) calculations we perform, mere construction of the rate matrix with order \(10^{17}\) elements would have been utterly impossible, not to mention time-evolution with that object.
  It is remarkable that the tensor network approach offers a way to evolve distributions rather than trajectories even in the face of that combinatorial explosion.

  \textit{Acknowledgment}.---We gratefully acknowledge Schuyler Nicholson and Phillip Helms for many insightful discussions.
  We are also grateful to Miles Stoudenmire, Matthew Fishman, Steven White, and other developers of ITensor, a library for implementing tensor network calculations, upon which this work was built.
 The material presented in this manuscript is based upon work supported by the National Science Foundation under Grant No.\ 2141385.

 \bibliography{biblio.bib}

\begin{thebibliography}{58}%
\makeatletter
\providecommand \@ifxundefined [1]{%
 \@ifx{#1\undefined}
}%
\providecommand \@ifnum [1]{%
 \ifnum #1\expandafter \@firstoftwo
 \else \expandafter \@secondoftwo
 \fi
}%
\providecommand \@ifx [1]{%
 \ifx #1\expandafter \@firstoftwo
 \else \expandafter \@secondoftwo
 \fi
}%
\providecommand \natexlab [1]{#1}%
\providecommand \enquote  [1]{``#1''}%
\providecommand \bibnamefont  [1]{#1}%
\providecommand \bibfnamefont [1]{#1}%
\providecommand \citenamefont [1]{#1}%
\providecommand \href@noop [0]{\@secondoftwo}%
\providecommand \href [0]{\begingroup \@sanitize@url \@href}%
\providecommand \@href[1]{\@@startlink{#1}\@@href}%
\providecommand \@@href[1]{\endgroup#1\@@endlink}%
\providecommand \@sanitize@url [0]{\catcode `\\12\catcode `\$12\catcode
  `\&12\catcode `\#12\catcode `\^12\catcode `\_12\catcode `\%12\relax}%
\providecommand \@@startlink[1]{}%
\providecommand \@@endlink[0]{}%
\providecommand \url  [0]{\begingroup\@sanitize@url \@url }%
\providecommand \@url [1]{\endgroup\@href {#1}{\urlprefix }}%
\providecommand \urlprefix  [0]{URL }%
\providecommand \Eprint [0]{\href }%
\providecommand \doibase [0]{http://dx.doi.org/}%
\providecommand \selectlanguage [0]{\@gobble}%
\providecommand \bibinfo  [0]{\@secondoftwo}%
\providecommand \bibfield  [0]{\@secondoftwo}%
\providecommand \translation [1]{[#1]}%
\providecommand \BibitemOpen [0]{}%
\providecommand \bibitemStop [0]{}%
\providecommand \bibitemNoStop [0]{.\EOS\space}%
\providecommand \EOS [0]{\spacefactor3000\relax}%
\providecommand \BibitemShut  [1]{\csname bibitem#1\endcsname}%
\let\auto@bib@innerbib\@empty
\bibitem [{\citenamefont {Smoluchowski}(1912)}]{smoluchowski1927experimentell}%
  \BibitemOpen
  \bibfield  {author} {\bibinfo {author} {\bibfnamefont {M.~V.}\ \bibnamefont
  {Smoluchowski}},\ }\href@noop {} {\bibfield  {journal} {\bibinfo  {journal}
  {Phys. Z.}\ }\textbf {\bibinfo {volume} {13}},\ \bibinfo {pages} {1069}
  (\bibinfo {year} {1912})}\BibitemShut {NoStop}%
\bibitem [{\citenamefont {Feynman}\ \emph {et~al.}(2015)\citenamefont
  {Feynman}, \citenamefont {Leighton},\ and\ \citenamefont
  {Sands}}]{feynman2011feynman}%
  \BibitemOpen
  \bibfield  {author} {\bibinfo {author} {\bibfnamefont {R.~P.}\ \bibnamefont
  {Feynman}}, \bibinfo {author} {\bibfnamefont {R.~B.}\ \bibnamefont
  {Leighton}}, \ and\ \bibinfo {author} {\bibfnamefont {M.}~\bibnamefont
  {Sands}},\ }\href@noop {} {\emph {\bibinfo {title} {The Feynman Lectures on
  Physics: The New Millennium Edition: Mainly Mechanics, Radiation, and
  Heat}}},\ Vol.~\bibinfo {volume} {1}\ (\bibinfo  {publisher} {Basic Books,
  New York, NY},\ \bibinfo {year} {2015})\ Chap.~\bibinfo {chapter}
  {46}\BibitemShut {NoStop}%
\bibitem [{\citenamefont {Harmer}\ \emph {et~al.}(2001)\citenamefont {Harmer},
  \citenamefont {Abbott}, \citenamefont {Taylor},\ and\ \citenamefont
  {Parrondo}}]{harmer2001brownian}%
  \BibitemOpen
  \bibfield  {author} {\bibinfo {author} {\bibfnamefont {G.~P.}\ \bibnamefont
  {Harmer}}, \bibinfo {author} {\bibfnamefont {D.}~\bibnamefont {Abbott}},
  \bibinfo {author} {\bibfnamefont {P.~G.}\ \bibnamefont {Taylor}}, \ and\
  \bibinfo {author} {\bibfnamefont {J.~M.}\ \bibnamefont {Parrondo}},\ }\href
  {\doibase 10.1063/1.1395623} {\bibfield  {journal} {\bibinfo  {journal}
  {Chaos}\ }\textbf {\bibinfo {volume} {11}},\ \bibinfo {pages} {705} (\bibinfo
  {year} {2001})}\BibitemShut {NoStop}%
\bibitem [{\citenamefont {Reimann}(2002)}]{reimann2002brownian}%
  \BibitemOpen
  \bibfield  {author} {\bibinfo {author} {\bibfnamefont {P.}~\bibnamefont
  {Reimann}},\ }\href {\doibase 10.1016/S0370-1573(01)00081-3} {\bibfield
  {journal} {\bibinfo  {journal} {Phys. Rep.}\ }\textbf {\bibinfo {volume}
  {361}},\ \bibinfo {pages} {57} (\bibinfo {year} {2002})}\BibitemShut
  {NoStop}%
\bibitem [{\citenamefont {Bartussek}\ \emph {et~al.}(1994)\citenamefont
  {Bartussek}, \citenamefont {H{\"a}nggi},\ and\ \citenamefont
  {Kissner}}]{bartussek1994periodically}%
  \BibitemOpen
  \bibfield  {author} {\bibinfo {author} {\bibfnamefont {R.}~\bibnamefont
  {Bartussek}}, \bibinfo {author} {\bibfnamefont {P.}~\bibnamefont
  {H{\"a}nggi}}, \ and\ \bibinfo {author} {\bibfnamefont {J.~G.}\ \bibnamefont
  {Kissner}},\ }\href {\doibase 10.1209/0295-5075/28/7/001} {\bibfield
  {journal} {\bibinfo  {journal} {Europhys. Lett.}\ }\textbf {\bibinfo {volume}
  {28}},\ \bibinfo {pages} {459} (\bibinfo {year} {1994})}\BibitemShut
  {NoStop}%
\bibitem [{\citenamefont {Elston}\ and\ \citenamefont
  {Doering}(1996)}]{elston1996numerical}%
  \BibitemOpen
  \bibfield  {author} {\bibinfo {author} {\bibfnamefont {T.~C.}\ \bibnamefont
  {Elston}}\ and\ \bibinfo {author} {\bibfnamefont {C.~R.}\ \bibnamefont
  {Doering}},\ }\href {\doibase 10.1007/BF02183737} {\bibfield  {journal}
  {\bibinfo  {journal} {J. Stat. Phys.}\ }\textbf {\bibinfo {volume} {83}},\
  \bibinfo {pages} {359–383} (\bibinfo {year} {1996})}\BibitemShut {NoStop}%
\bibitem [{\citenamefont {Reimann}\ and\ \citenamefont
  {H{\"a}nggi}(2002)}]{reimann2001introduction}%
  \BibitemOpen
  \bibfield  {author} {\bibinfo {author} {\bibfnamefont {P.}~\bibnamefont
  {Reimann}}\ and\ \bibinfo {author} {\bibfnamefont {P.}~\bibnamefont
  {H{\"a}nggi}},\ }\href {\doibase 10.1007/s003390201331} {\bibfield  {journal}
  {\bibinfo  {journal} {Appl. Phys. A}\ }\textbf {\bibinfo {volume} {75}},\
  \bibinfo {pages} {169} (\bibinfo {year} {2002})}\BibitemShut {NoStop}%
\bibitem [{\citenamefont {B{\"u}ttiker}(1987)}]{buttiker1987transport}%
  \BibitemOpen
  \bibfield  {author} {\bibinfo {author} {\bibfnamefont {M.}~\bibnamefont
  {B{\"u}ttiker}},\ }\href {\doibase 10.1007/BF01304221} {\bibfield  {journal}
  {\bibinfo  {journal} {Z. Phys. B, Cond. Matter}\ }\textbf {\bibinfo {volume}
  {68}},\ \bibinfo {pages} {161} (\bibinfo {year} {1987})}\BibitemShut
  {NoStop}%
\bibitem [{\citenamefont {Kostur}\ and\ \citenamefont
  {\L{}uczka}(2001)}]{kostur2001multiple}%
  \BibitemOpen
  \bibfield  {author} {\bibinfo {author} {\bibfnamefont {M.}~\bibnamefont
  {Kostur}}\ and\ \bibinfo {author} {\bibfnamefont {J.}~\bibnamefont
  {\L{}uczka}},\ }\href {\doibase 10.1103/PhysRevE.63.021101} {\bibfield
  {journal} {\bibinfo  {journal} {Phys. Rev. E}\ }\textbf {\bibinfo {volume}
  {63}},\ \bibinfo {pages} {021101} (\bibinfo {year} {2001})}\BibitemShut
  {NoStop}%
\bibitem [{\citenamefont {Tammelo}\ \emph {et~al.}(2002)\citenamefont
  {Tammelo}, \citenamefont {Mankin},\ and\ \citenamefont
  {Martila}}]{tammelo2002three}%
  \BibitemOpen
  \bibfield  {author} {\bibinfo {author} {\bibfnamefont {R.}~\bibnamefont
  {Tammelo}}, \bibinfo {author} {\bibfnamefont {R.}~\bibnamefont {Mankin}}, \
  and\ \bibinfo {author} {\bibfnamefont {D.}~\bibnamefont {Martila}},\ }\href
  {\doibase 10.1103/PhysRevE.66.051101} {\bibfield  {journal} {\bibinfo
  {journal} {Phys. Rev. E}\ }\textbf {\bibinfo {volume} {66}},\ \bibinfo
  {pages} {051101} (\bibinfo {year} {2002})}\BibitemShut {NoStop}%
\bibitem [{\citenamefont {Zeng}\ \emph {et~al.}(2010)\citenamefont {Zeng},
  \citenamefont {Gong},\ and\ \citenamefont {Tian}}]{zeng2010current}%
  \BibitemOpen
  \bibfield  {author} {\bibinfo {author} {\bibfnamefont {C.}~\bibnamefont
  {Zeng}}, \bibinfo {author} {\bibfnamefont {A.}~\bibnamefont {Gong}}, \ and\
  \bibinfo {author} {\bibfnamefont {Y.}~\bibnamefont {Tian}},\ }\href {\doibase
  10.1016/j.physa.2009.12.059} {\bibfield  {journal} {\bibinfo  {journal}
  {Physica A}\ }\textbf {\bibinfo {volume} {389}},\ \bibinfo {pages} {1971}
  (\bibinfo {year} {2010})}\BibitemShut {NoStop}%
\bibitem [{\citenamefont {Kedem}\ and\ \citenamefont
  {Weiss}(2019)}]{kedem2019cooperative}%
  \BibitemOpen
  \bibfield  {author} {\bibinfo {author} {\bibfnamefont {O.}~\bibnamefont
  {Kedem}}\ and\ \bibinfo {author} {\bibfnamefont {E.~A.}\ \bibnamefont
  {Weiss}},\ }\href {\doibase 10.1021/acs.jpcc.9b00344} {\bibfield  {journal}
  {\bibinfo  {journal} {J. Phys. Chem. C}\ }\textbf {\bibinfo {volume} {123}},\
  \bibinfo {pages} {6913} (\bibinfo {year} {2019})}\BibitemShut {NoStop}%
\bibitem [{\citenamefont {Kodaimati}\ \emph {et~al.}(2019)\citenamefont
  {Kodaimati}, \citenamefont {Kedem}, \citenamefont {Schatz},\ and\
  \citenamefont {Weiss}}]{kodaimati2019empirical}%
  \BibitemOpen
  \bibfield  {author} {\bibinfo {author} {\bibfnamefont {M.~S.}\ \bibnamefont
  {Kodaimati}}, \bibinfo {author} {\bibfnamefont {O.}~\bibnamefont {Kedem}},
  \bibinfo {author} {\bibfnamefont {G.~C.}\ \bibnamefont {Schatz}}, \ and\
  \bibinfo {author} {\bibfnamefont {E.~A.}\ \bibnamefont {Weiss}},\ }\href
  {\doibase 10.1021/acs.jpcc.9b06503} {\bibfield  {journal} {\bibinfo
  {journal} {J. Phys. Chem. C}\ }\textbf {\bibinfo {volume} {123}},\ \bibinfo
  {pages} {22050} (\bibinfo {year} {2019})}\BibitemShut {NoStop}%
\bibitem [{\citenamefont {Lau}\ and\ \citenamefont
  {Kedem}(2020)}]{lau2020electron}%
  \BibitemOpen
  \bibfield  {author} {\bibinfo {author} {\bibfnamefont {B.}~\bibnamefont
  {Lau}}\ and\ \bibinfo {author} {\bibfnamefont {O.}~\bibnamefont {Kedem}},\
  }\href {\doibase 10.1063/5.0009561} {\bibfield  {journal} {\bibinfo
  {journal} {J. Chem. Phys.}\ }\textbf {\bibinfo {volume} {152}},\ \bibinfo
  {pages} {200901} (\bibinfo {year} {2020})}\BibitemShut {NoStop}%
\bibitem [{\citenamefont {Craig}\ \emph {et~al.}(2006)\citenamefont {Craig},
  \citenamefont {Zuckermann},\ and\ \citenamefont {Linke}}]{Craig2006}%
  \BibitemOpen
  \bibfield  {author} {\bibinfo {author} {\bibfnamefont {E.~M.}\ \bibnamefont
  {Craig}}, \bibinfo {author} {\bibfnamefont {M.~J.}\ \bibnamefont
  {Zuckermann}}, \ and\ \bibinfo {author} {\bibfnamefont {H.}~\bibnamefont
  {Linke}},\ }\href {\doibase 10.1103/PhysRevE.73.051106} {\bibfield  {journal}
  {\bibinfo  {journal} {Phys. Rev. E}\ }\textbf {\bibinfo {volume} {73}},\
  \bibinfo {pages} {051106} (\bibinfo {year} {2006})}\BibitemShut {NoStop}%
\bibitem [{\citenamefont {Li}\ \emph {et~al.}(2016)\citenamefont {Li},
  \citenamefont {Chen},\ and\ \citenamefont {Zheng}}]{Li2016a}%
  \BibitemOpen
  \bibfield  {author} {\bibinfo {author} {\bibfnamefont {C.-P.}\ \bibnamefont
  {Li}}, \bibinfo {author} {\bibfnamefont {H.-B.}\ \bibnamefont {Chen}}, \ and\
  \bibinfo {author} {\bibfnamefont {Z.-G.}\ \bibnamefont {Zheng}},\ }\href
  {\doibase 10.1007/s11467-016-0622-1} {\bibfield  {journal} {\bibinfo
  {journal} {Front. Phys.}\ }\textbf {\bibinfo {volume} {12}},\ \bibinfo
  {pages} {120502} (\bibinfo {year} {2016})}\BibitemShut {NoStop}%
\bibitem [{\citenamefont {Klumpp}\ \emph {et~al.}(2005)\citenamefont {Klumpp},
  \citenamefont {Nieuwenhuizen},\ and\ \citenamefont {Lipowsky}}]{Klumpp2005}%
  \BibitemOpen
  \bibfield  {author} {\bibinfo {author} {\bibfnamefont {S.}~\bibnamefont
  {Klumpp}}, \bibinfo {author} {\bibfnamefont {T.~M.}\ \bibnamefont
  {Nieuwenhuizen}}, \ and\ \bibinfo {author} {\bibfnamefont {R.}~\bibnamefont
  {Lipowsky}},\ }\href {\doibase 10.1016/j.physe.2005.05.037} {\bibfield
  {journal} {\bibinfo  {journal} {Physica E Low Dimens. Syst. Nanostruct.}\
  }\bibinfo {series} {Frontiers of {{Quantum}}},\ \textbf {\bibinfo {volume}
  {29}},\ \bibinfo {pages} {380} (\bibinfo {year} {2005})}\BibitemShut
  {NoStop}%
\bibitem [{\citenamefont {Imparato}(2015)}]{Imparato2015}%
  \BibitemOpen
  \bibfield  {author} {\bibinfo {author} {\bibfnamefont {A.}~\bibnamefont
  {Imparato}},\ }\href {\doibase 10.1088/1367-2630/17/12/125004} {\bibfield
  {journal} {\bibinfo  {journal} {New J. Phys.}\ }\textbf {\bibinfo {volume}
  {17}},\ \bibinfo {pages} {125004} (\bibinfo {year} {2015})}\BibitemShut
  {NoStop}%
\bibitem [{\citenamefont {Golubeva}\ and\ \citenamefont
  {Imparato}(2012)}]{Golubeva2012}%
  \BibitemOpen
  \bibfield  {author} {\bibinfo {author} {\bibfnamefont {N.}~\bibnamefont
  {Golubeva}}\ and\ \bibinfo {author} {\bibfnamefont {A.}~\bibnamefont
  {Imparato}},\ }\href {\doibase 10.1103/PhysRevLett.109.190602} {\bibfield
  {journal} {\bibinfo  {journal} {Phys. Rev. Lett.}\ }\textbf {\bibinfo
  {volume} {109}},\ \bibinfo {pages} {190602} (\bibinfo {year}
  {2012})}\BibitemShut {NoStop}%
\bibitem [{\citenamefont {Camp{\`a}s}\ \emph {et~al.}(2006)\citenamefont
  {Camp{\`a}s}, \citenamefont {Kafri}, \citenamefont {Zeldovich}, \citenamefont
  {Casademunt},\ and\ \citenamefont {Joanny}}]{Campas2006}%
  \BibitemOpen
  \bibfield  {author} {\bibinfo {author} {\bibfnamefont {O.}~\bibnamefont
  {Camp{\`a}s}}, \bibinfo {author} {\bibfnamefont {Y.}~\bibnamefont {Kafri}},
  \bibinfo {author} {\bibfnamefont {K.~B.}\ \bibnamefont {Zeldovich}}, \bibinfo
  {author} {\bibfnamefont {J.}~\bibnamefont {Casademunt}}, \ and\ \bibinfo
  {author} {\bibfnamefont {J.-F.}\ \bibnamefont {Joanny}},\ }\href {\doibase
  10.1103/PhysRevLett.97.038101} {\bibfield  {journal} {\bibinfo  {journal}
  {Phys. Rev. Lett.}\ }\textbf {\bibinfo {volume} {97}},\ \bibinfo {pages}
  {038101} (\bibinfo {year} {2006})}\BibitemShut {NoStop}%
\bibitem [{\citenamefont {Vroylandt}\ \emph {et~al.}(2020)\citenamefont
  {Vroylandt}, \citenamefont {Esposito},\ and\ \citenamefont
  {Verley}}]{Vroylandt2020b}%
  \BibitemOpen
  \bibfield  {author} {\bibinfo {author} {\bibfnamefont {H.}~\bibnamefont
  {Vroylandt}}, \bibinfo {author} {\bibfnamefont {M.}~\bibnamefont {Esposito}},
  \ and\ \bibinfo {author} {\bibfnamefont {G.}~\bibnamefont {Verley}},\ }\href
  {\doibase 10.1103/PhysRevLett.124.250603} {\bibfield  {journal} {\bibinfo
  {journal} {Phys. Rev. Lett.}\ }\textbf {\bibinfo {volume} {124}},\ \bibinfo
  {pages} {250603} (\bibinfo {year} {2020})}\BibitemShut {NoStop}%
\bibitem [{\citenamefont {Chou}\ \emph {et~al.}(2011)\citenamefont {Chou},
  \citenamefont {Mallick},\ and\ \citenamefont {Zia}}]{chou2011non}%
  \BibitemOpen
  \bibfield  {author} {\bibinfo {author} {\bibfnamefont {T.}~\bibnamefont
  {Chou}}, \bibinfo {author} {\bibfnamefont {K.}~\bibnamefont {Mallick}}, \
  and\ \bibinfo {author} {\bibfnamefont {R.}~\bibnamefont {Zia}},\ }\href
  {\doibase 10.1088/0034-4885/74/11/116601} {\bibfield  {journal} {\bibinfo
  {journal} {Rep. Prog. Phys.}\ }\textbf {\bibinfo {volume} {74}},\ \bibinfo
  {pages} {116601} (\bibinfo {year} {2011})}\BibitemShut {NoStop}%
\bibitem [{\citenamefont {Lazarescu}(2015)}]{lazarescu2015physicist}%
  \BibitemOpen
  \bibfield  {author} {\bibinfo {author} {\bibfnamefont {A.}~\bibnamefont
  {Lazarescu}},\ }\href {\doibase 10.1088/1751-8113/48/50/503001} {\bibfield
  {journal} {\bibinfo  {journal} {J. Phys. A}\ }\textbf {\bibinfo {volume}
  {48}},\ \bibinfo {pages} {503001} (\bibinfo {year} {2015})}\BibitemShut
  {NoStop}%
\bibitem [{\citenamefont {Helms}\ \emph {et~al.}(2019)\citenamefont {Helms},
  \citenamefont {Ray},\ and\ \citenamefont {Chan}}]{helms2019dynamical}%
  \BibitemOpen
  \bibfield  {author} {\bibinfo {author} {\bibfnamefont {P.}~\bibnamefont
  {Helms}}, \bibinfo {author} {\bibfnamefont {U.}~\bibnamefont {Ray}}, \ and\
  \bibinfo {author} {\bibfnamefont {G.~K.-L.}\ \bibnamefont {Chan}},\ }\href
  {\doibase 10.1103/PhysRevE.100.022101} {\bibfield  {journal} {\bibinfo
  {journal} {Phys. Rev. E}\ }\textbf {\bibinfo {volume} {100}},\ \bibinfo
  {pages} {022101} (\bibinfo {year} {2019})}\BibitemShut {NoStop}%
\bibitem [{\citenamefont {Helms}\ and\ \citenamefont
  {Chan}(2020)}]{helms2020dynamical}%
  \BibitemOpen
  \bibfield  {author} {\bibinfo {author} {\bibfnamefont {P.}~\bibnamefont
  {Helms}}\ and\ \bibinfo {author} {\bibfnamefont {G.~K.-L.}\ \bibnamefont
  {Chan}},\ }\href {\doibase 10.1103/PhysRevLett.125.140601} {\bibfield
  {journal} {\bibinfo  {journal} {Phys. Rev. Lett.}\ }\textbf {\bibinfo
  {volume} {125}},\ \bibinfo {pages} {140601} (\bibinfo {year}
  {2020})}\BibitemShut {NoStop}%
\bibitem [{\citenamefont {Proeme}\ \emph {et~al.}(2010)\citenamefont {Proeme},
  \citenamefont {Blythe},\ and\ \citenamefont {Evans}}]{Proeme2010}%
  \BibitemOpen
  \bibfield  {author} {\bibinfo {author} {\bibfnamefont {A.}~\bibnamefont
  {Proeme}}, \bibinfo {author} {\bibfnamefont {R.~A.}\ \bibnamefont {Blythe}},
  \ and\ \bibinfo {author} {\bibfnamefont {M.~R.}\ \bibnamefont {Evans}},\
  }\href {\doibase 10.1088/1751-8113/44/3/035003} {\bibfield  {journal}
  {\bibinfo  {journal} {J. Phys. A Math. Theor.}\ }\textbf {\bibinfo {volume}
  {44}},\ \bibinfo {pages} {035003} (\bibinfo {year} {2010})}\BibitemShut
  {NoStop}%
\bibitem [{\citenamefont {Strand}\ \emph {et~al.}(2020)\citenamefont {Strand},
  \citenamefont {Fu},\ and\ \citenamefont {Gingrich}}]{strand2020current}%
  \BibitemOpen
  \bibfield  {author} {\bibinfo {author} {\bibfnamefont {N.~E.}\ \bibnamefont
  {Strand}}, \bibinfo {author} {\bibfnamefont {R.-S.}\ \bibnamefont {Fu}}, \
  and\ \bibinfo {author} {\bibfnamefont {T.~R.}\ \bibnamefont {Gingrich}},\
  }\href {\doibase 10.1103/PhysRevE.102.012141} {\bibfield  {journal} {\bibinfo
   {journal} {Phys. Rev. E}\ }\textbf {\bibinfo {volume} {102}},\ \bibinfo
  {pages} {012141} (\bibinfo {year} {2020})}\BibitemShut {NoStop}%
\bibitem [{\citenamefont {Schollw{\"o}ck}(2011)}]{schollwock2011density}%
  \BibitemOpen
  \bibfield  {author} {\bibinfo {author} {\bibfnamefont {U.}~\bibnamefont
  {Schollw{\"o}ck}},\ }\href {\doibase 10.1016/j.aop.2010.09.012} {\bibfield
  {journal} {\bibinfo  {journal} {Ann. Physics}\ }\textbf {\bibinfo {volume}
  {326}},\ \bibinfo {pages} {96} (\bibinfo {year} {2011})}\BibitemShut
  {NoStop}%
\bibitem [{\citenamefont {Paeckel}\ \emph {et~al.}(2019)\citenamefont
  {Paeckel}, \citenamefont {K{\"o}hler}, \citenamefont {Swoboda}, \citenamefont
  {Manmana}, \citenamefont {Schollw{\"o}ck},\ and\ \citenamefont
  {Hubig}}]{paeckel2019time}%
  \BibitemOpen
  \bibfield  {author} {\bibinfo {author} {\bibfnamefont {S.}~\bibnamefont
  {Paeckel}}, \bibinfo {author} {\bibfnamefont {T.}~\bibnamefont {K{\"o}hler}},
  \bibinfo {author} {\bibfnamefont {A.}~\bibnamefont {Swoboda}}, \bibinfo
  {author} {\bibfnamefont {S.~R.}\ \bibnamefont {Manmana}}, \bibinfo {author}
  {\bibfnamefont {U.}~\bibnamefont {Schollw{\"o}ck}}, \ and\ \bibinfo {author}
  {\bibfnamefont {C.}~\bibnamefont {Hubig}},\ }\href {\doibase
  10.1016/j.aop.2019.167998} {\bibfield  {journal} {\bibinfo  {journal} {Ann.
  Physics}\ }\textbf {\bibinfo {volume} {411}},\ \bibinfo {pages} {167998}
  (\bibinfo {year} {2019})}\BibitemShut {NoStop}%
\bibitem [{\citenamefont {Ba{\~n}uls}\ and\ \citenamefont
  {Garrahan}(2019)}]{banuls2019using}%
  \BibitemOpen
  \bibfield  {author} {\bibinfo {author} {\bibfnamefont {M.~C.}\ \bibnamefont
  {Ba{\~n}uls}}\ and\ \bibinfo {author} {\bibfnamefont {J.~P.}\ \bibnamefont
  {Garrahan}},\ }\href {\doibase 10.1103/PhysRevLett.123.200601} {\bibfield
  {journal} {\bibinfo  {journal} {Phys. Rev. Lett.}\ }\textbf {\bibinfo
  {volume} {123}},\ \bibinfo {pages} {200601} (\bibinfo {year}
  {2019})}\BibitemShut {NoStop}%
\bibitem [{\citenamefont {Nagy}\ \emph {et~al.}(2002)\citenamefont {Nagy},
  \citenamefont {Appert},\ and\ \citenamefont {Santen}}]{Nagy2002}%
  \BibitemOpen
  \bibfield  {author} {\bibinfo {author} {\bibfnamefont {Z.}~\bibnamefont
  {Nagy}}, \bibinfo {author} {\bibfnamefont {C.}~\bibnamefont {Appert}}, \ and\
  \bibinfo {author} {\bibfnamefont {L.}~\bibnamefont {Santen}},\ }\href
  {\doibase 10.1023/A:1020462531383} {\bibfield  {journal} {\bibinfo  {journal}
  {J. Stat. Phys.}\ }\textbf {\bibinfo {volume} {109}},\ \bibinfo {pages} {623}
  (\bibinfo {year} {2002})}\BibitemShut {NoStop}%
\bibitem [{\citenamefont {Hieida}(1998)}]{Hieida1998}%
  \BibitemOpen
  \bibfield  {author} {\bibinfo {author} {\bibfnamefont {Y.}~\bibnamefont
  {Hieida}},\ }\href {\doibase 10.1143/JPSJ.67.369} {\bibfield  {journal}
  {\bibinfo  {journal} {J. Phys. Soc. Japan}\ }\textbf {\bibinfo {volume}
  {67}},\ \bibinfo {pages} {369} (\bibinfo {year} {1998})}\BibitemShut
  {NoStop}%
\bibitem [{\citenamefont {Temme}\ and\ \citenamefont
  {Verstraete}(2010)}]{Temme2010}%
  \BibitemOpen
  \bibfield  {author} {\bibinfo {author} {\bibfnamefont {K.}~\bibnamefont
  {Temme}}\ and\ \bibinfo {author} {\bibfnamefont {F.}~\bibnamefont
  {Verstraete}},\ }\href {\doibase 10.1103/PhysRevLett.104.210502} {\bibfield
  {journal} {\bibinfo  {journal} {Phys. Rev. Lett.}\ }\textbf {\bibinfo
  {volume} {104}},\ \bibinfo {pages} {210502} (\bibinfo {year}
  {2010})}\BibitemShut {NoStop}%
\bibitem [{\citenamefont {Johnson}\ \emph {et~al.}(2010)\citenamefont
  {Johnson}, \citenamefont {Clark},\ and\ \citenamefont
  {Jaksch}}]{Johnson2010}%
  \BibitemOpen
  \bibfield  {author} {\bibinfo {author} {\bibfnamefont {T.~H.}\ \bibnamefont
  {Johnson}}, \bibinfo {author} {\bibfnamefont {S.~R.}\ \bibnamefont {Clark}},
  \ and\ \bibinfo {author} {\bibfnamefont {D.}~\bibnamefont {Jaksch}},\ }\href
  {\doibase 10.1103/PhysRevE.82.036702} {\bibfield  {journal} {\bibinfo
  {journal} {Phys. Rev. E}\ }\textbf {\bibinfo {volume} {82}},\ \bibinfo
  {pages} {036702} (\bibinfo {year} {2010})}\BibitemShut {NoStop}%
\bibitem [{\citenamefont {Johnson}\ \emph {et~al.}(2015)\citenamefont
  {Johnson}, \citenamefont {Elliott}, \citenamefont {Clark},\ and\
  \citenamefont {Jaksch}}]{Johnson2015}%
  \BibitemOpen
  \bibfield  {author} {\bibinfo {author} {\bibfnamefont {T.~H.}\ \bibnamefont
  {Johnson}}, \bibinfo {author} {\bibfnamefont {T.~J.}\ \bibnamefont
  {Elliott}}, \bibinfo {author} {\bibfnamefont {S.~R.}\ \bibnamefont {Clark}},
  \ and\ \bibinfo {author} {\bibfnamefont {D.}~\bibnamefont {Jaksch}},\ }\href
  {\doibase 10.1103/PhysRevLett.114.090602} {\bibfield  {journal} {\bibinfo
  {journal} {Phys. Rev. Lett.}\ }\textbf {\bibinfo {volume} {114}},\ \bibinfo
  {pages} {090602} (\bibinfo {year} {2015})}\BibitemShut {NoStop}%
\bibitem [{\citenamefont {White}(1992)}]{white1992density}%
  \BibitemOpen
  \bibfield  {author} {\bibinfo {author} {\bibfnamefont {S.~R.}\ \bibnamefont
  {White}},\ }\href {\doibase 10.1103/PhysRevLett.69.2863} {\bibfield
  {journal} {\bibinfo  {journal} {Phys. Rev. Lett.}\ }\textbf {\bibinfo
  {volume} {69}},\ \bibinfo {pages} {2863} (\bibinfo {year}
  {1992})}\BibitemShut {NoStop}%
\bibitem [{\citenamefont {Haegeman}\ \emph {et~al.}(2011)\citenamefont
  {Haegeman}, \citenamefont {Cirac}, \citenamefont {Osborne}, \citenamefont
  {Pi{\v z}orn}, \citenamefont {Verschelde},\ and\ \citenamefont
  {Verstraete}}]{haegeman2011time}%
  \BibitemOpen
  \bibfield  {author} {\bibinfo {author} {\bibfnamefont {J.}~\bibnamefont
  {Haegeman}}, \bibinfo {author} {\bibfnamefont {J.~I.}\ \bibnamefont {Cirac}},
  \bibinfo {author} {\bibfnamefont {T.~J.}\ \bibnamefont {Osborne}}, \bibinfo
  {author} {\bibfnamefont {I.}~\bibnamefont {Pi{\v z}orn}}, \bibinfo {author}
  {\bibfnamefont {H.}~\bibnamefont {Verschelde}}, \ and\ \bibinfo {author}
  {\bibfnamefont {F.}~\bibnamefont {Verstraete}},\ }\href {\doibase
  10.1103/PhysRevLett.107.070601} {\bibfield  {journal} {\bibinfo  {journal}
  {Phys. Rev. Lett.}\ }\textbf {\bibinfo {volume} {107}},\ \bibinfo {pages}
  {070601} (\bibinfo {year} {2011})}\BibitemShut {NoStop}%
\bibitem [{\citenamefont {Haegeman}\ \emph {et~al.}(2016)\citenamefont
  {Haegeman}, \citenamefont {Lubich}, \citenamefont {Oseledets}, \citenamefont
  {Vandereycken},\ and\ \citenamefont {Verstraete}}]{haegeman2016unifying}%
  \BibitemOpen
  \bibfield  {author} {\bibinfo {author} {\bibfnamefont {J.}~\bibnamefont
  {Haegeman}}, \bibinfo {author} {\bibfnamefont {C.}~\bibnamefont {Lubich}},
  \bibinfo {author} {\bibfnamefont {I.}~\bibnamefont {Oseledets}}, \bibinfo
  {author} {\bibfnamefont {B.}~\bibnamefont {Vandereycken}}, \ and\ \bibinfo
  {author} {\bibfnamefont {F.}~\bibnamefont {Verstraete}},\ }\href {\doibase
  10.1103/PhysRevB.94.165116} {\bibfield  {journal} {\bibinfo  {journal} {Phys.
  Rev. B}\ }\textbf {\bibinfo {volume} {94}},\ \bibinfo {pages} {165116}
  (\bibinfo {year} {2016})}\BibitemShut {NoStop}%
\bibitem [{\citenamefont {Bauernfeind}\ and\ \citenamefont
  {Aichhorn}(2020)}]{bauernfeind2020time}%
  \BibitemOpen
  \bibfield  {author} {\bibinfo {author} {\bibfnamefont {D.}~\bibnamefont
  {Bauernfeind}}\ and\ \bibinfo {author} {\bibfnamefont {M.}~\bibnamefont
  {Aichhorn}},\ }\href {\doibase 10.21468/SciPostPhys.8.2.024} {\bibfield
  {journal} {\bibinfo  {journal} {SciPost Phys.}\ }\textbf {\bibinfo {volume}
  {8}},\ \bibinfo {pages} {024} (\bibinfo {year} {2020})}\BibitemShut {NoStop}%
\bibitem [{\citenamefont {Yang}\ and\ \citenamefont {White}(2020)}]{Yang2020a}%
  \BibitemOpen
  \bibfield  {author} {\bibinfo {author} {\bibfnamefont {M.}~\bibnamefont
  {Yang}}\ and\ \bibinfo {author} {\bibfnamefont {S.~R.}\ \bibnamefont
  {White}},\ }\href {\doibase 10.1103/PhysRevB.102.094315} {\bibfield
  {journal} {\bibinfo  {journal} {Phys. Rev. B}\ }\textbf {\bibinfo {volume}
  {102}},\ \bibinfo {pages} {094315} (\bibinfo {year} {2020})}\BibitemShut
  {NoStop}%
\bibitem [{\citenamefont {Kloss}\ \emph {et~al.}(2018)\citenamefont {Kloss},
  \citenamefont {Lev},\ and\ \citenamefont {Reichman}}]{kloss2018time}%
  \BibitemOpen
  \bibfield  {author} {\bibinfo {author} {\bibfnamefont {B.}~\bibnamefont
  {Kloss}}, \bibinfo {author} {\bibfnamefont {Y.~B.}\ \bibnamefont {Lev}}, \
  and\ \bibinfo {author} {\bibfnamefont {D.}~\bibnamefont {Reichman}},\ }\href
  {\doibase 10.1103/PhysRevB.97.024307} {\bibfield  {journal} {\bibinfo
  {journal} {Phys. Rev. B}\ }\textbf {\bibinfo {volume} {97}},\ \bibinfo
  {pages} {024307} (\bibinfo {year} {2018})}\BibitemShut {NoStop}%
\bibitem [{\citenamefont {Gorissen}\ and\ \citenamefont
  {Vanderzande}(2012)}]{Gorissen2012}%
  \BibitemOpen
  \bibfield  {author} {\bibinfo {author} {\bibfnamefont {M.}~\bibnamefont
  {Gorissen}}\ and\ \bibinfo {author} {\bibfnamefont {C.}~\bibnamefont
  {Vanderzande}},\ }\href {\doibase 10.1103/PhysRevE.86.051114} {\bibfield
  {journal} {\bibinfo  {journal} {Phys. Rev. E}\ }\textbf {\bibinfo {volume}
  {86}},\ \bibinfo {pages} {051114} (\bibinfo {year} {2012})}\BibitemShut
  {NoStop}%
\bibitem [{\citenamefont {Gorissen}\ and\ \citenamefont
  {Vanderzande}(2011)}]{Gorissen2011}%
  \BibitemOpen
  \bibfield  {author} {\bibinfo {author} {\bibfnamefont {M.}~\bibnamefont
  {Gorissen}}\ and\ \bibinfo {author} {\bibfnamefont {C.}~\bibnamefont
  {Vanderzande}},\ }\href {\doibase 10.1088/1751-8113/44/11/115005} {\bibfield
  {journal} {\bibinfo  {journal} {J. Phys. A Math. Theor.}\ }\textbf {\bibinfo
  {volume} {44}},\ \bibinfo {pages} {115005} (\bibinfo {year}
  {2011})}\BibitemShut {NoStop}%
\bibitem [{\citenamefont {Lazarescu}(2013)}]{Lazarescu2013}%
  \BibitemOpen
  \bibfield  {author} {\bibinfo {author} {\bibfnamefont {A.}~\bibnamefont
  {Lazarescu}},\ }\href {\doibase 10.1088/1751-8113/46/14/145003} {\bibfield
  {journal} {\bibinfo  {journal} {J. Phys. A Math. Theor.}\ }\textbf {\bibinfo
  {volume} {46}},\ \bibinfo {pages} {145003} (\bibinfo {year}
  {2013})}\BibitemShut {NoStop}%
\bibitem [{\citenamefont {Causer}\ \emph {et~al.}(2020)\citenamefont {Causer},
  , \citenamefont {Ba{\~n}uls},\ and\ \citenamefont
  {Garrahan}}]{Causer2021optimal}%
  \BibitemOpen
  \bibfield  {author} {\bibinfo {author} {\bibfnamefont {L.}~\bibnamefont
  {Causer}}, , \bibinfo {author} {\bibfnamefont {M.~C.}\ \bibnamefont
  {Ba{\~n}uls}}, \ and\ \bibinfo {author} {\bibfnamefont {J.~P.}\ \bibnamefont
  {Garrahan}},\ }\href {\doibase 10.1103/PhysRevE.103.062144} {\bibfield
  {journal} {\bibinfo  {journal} {Phys. Rev. E}\ }\textbf {\bibinfo {volume}
  {103}},\ \bibinfo {pages} {062144} (\bibinfo {year} {2020})}\BibitemShut
  {NoStop}%
\bibitem [{\citenamefont {Causer}\ \emph {et~al.}(2021)\citenamefont {Causer},
  \citenamefont {Ba{\~n}uls},\ and\ \citenamefont
  {Garrahan}}]{Causer2021finite}%
  \BibitemOpen
  \bibfield  {author} {\bibinfo {author} {\bibfnamefont {L.}~\bibnamefont
  {Causer}}, \bibinfo {author} {\bibfnamefont {M.~C.}\ \bibnamefont
  {Ba{\~n}uls}}, \ and\ \bibinfo {author} {\bibfnamefont {J.~P.}\ \bibnamefont
  {Garrahan}},\ }\href@noop {} {\bibfield  {journal} {\bibinfo  {journal}
  {arXiv:2108.11418}\ } (\bibinfo {year} {2021})},\ \Eprint
  {http://arxiv.org/abs/2108.11418} {2108.11418} \BibitemShut {NoStop}%
\bibitem [{\citenamefont {Gingrich}\ \emph {et~al.}(2017)\citenamefont
  {Gingrich}, \citenamefont {Rotskoff},\ and\ \citenamefont
  {Horowitz}}]{gingrich2017inferring}%
  \BibitemOpen
  \bibfield  {author} {\bibinfo {author} {\bibfnamefont {T.~R.}\ \bibnamefont
  {Gingrich}}, \bibinfo {author} {\bibfnamefont {G.~M.}\ \bibnamefont
  {Rotskoff}}, \ and\ \bibinfo {author} {\bibfnamefont {J.~M.}\ \bibnamefont
  {Horowitz}},\ }\href {\doibase 10.1088/1751-8121/aa672f} {\bibfield
  {journal} {\bibinfo  {journal} {J. Phys. A Math. Theor.}\ }\textbf {\bibinfo
  {volume} {50}},\ \bibinfo {pages} {184004} (\bibinfo {year}
  {2017})}\BibitemShut {NoStop}%
\bibitem [{\citenamefont {Kedem}\ \emph {et~al.}(2017)\citenamefont {Kedem},
  \citenamefont {Lau}, \citenamefont {Ratner},\ and\ \citenamefont
  {Weiss}}]{kedem2017light}%
  \BibitemOpen
  \bibfield  {author} {\bibinfo {author} {\bibfnamefont {O.}~\bibnamefont
  {Kedem}}, \bibinfo {author} {\bibfnamefont {B.}~\bibnamefont {Lau}}, \bibinfo
  {author} {\bibfnamefont {M.~A.}\ \bibnamefont {Ratner}}, \ and\ \bibinfo
  {author} {\bibfnamefont {E.~A.}\ \bibnamefont {Weiss}},\ }\href {\doibase
  10.1073/pnas.1705973114} {\bibfield  {journal} {\bibinfo  {journal} {Proc.
  Natl. Acad. Sci. U.S.A.}\ }\textbf {\bibinfo {volume} {114}},\ \bibinfo
  {pages} {8698} (\bibinfo {year} {2017})}\BibitemShut {NoStop}%
\bibitem [{\citenamefont {van Kampen}(2007)}]{vankampen2007stochastic}%
  \BibitemOpen
  \bibfield  {author} {\bibinfo {author} {\bibfnamefont {N.}~\bibnamefont {van
  Kampen}},\ }\href@noop {} {\emph {\bibinfo {title} {Stochastic Processes in
  Physics and Chemistry}}}\ (\bibinfo  {publisher} {North Holland},\ \bibinfo
  {year} {2007})\BibitemShut {NoStop}%
\bibitem [{\citenamefont {Lebowitz}\ and\ \citenamefont
  {Spohn}(1999)}]{lebowitz1999}%
  \BibitemOpen
  \bibfield  {author} {\bibinfo {author} {\bibfnamefont {J.~L.}\ \bibnamefont
  {Lebowitz}}\ and\ \bibinfo {author} {\bibfnamefont {H.}~\bibnamefont
  {Spohn}},\ }\href {\doibase 10.1023/A:1004589714161} {\bibfield  {journal}
  {\bibinfo  {journal} {J. Stat. Phys.}\ }\textbf {\bibinfo {volume} {95}},\
  \bibinfo {pages} {333} (\bibinfo {year} {1999})}\BibitemShut {NoStop}%
\bibitem [{\citenamefont {Lecomte}\ \emph {et~al.}(2007)\citenamefont
  {Lecomte}, \citenamefont {Appert-Rolland},\ and\ \citenamefont {van
  Wijland}}]{Lecomte2007}%
  \BibitemOpen
  \bibfield  {author} {\bibinfo {author} {\bibfnamefont {V.}~\bibnamefont
  {Lecomte}}, \bibinfo {author} {\bibfnamefont {C.}~\bibnamefont
  {Appert-Rolland}}, \ and\ \bibinfo {author} {\bibfnamefont {F.}~\bibnamefont
  {van Wijland}},\ }\href {\doibase 10.1007/s10955-006-9254-0} {\bibfield
  {journal} {\bibinfo  {journal} {J. Stat. Phys.}\ }\textbf {\bibinfo {volume}
  {127}},\ \bibinfo {pages} {51} (\bibinfo {year} {2007})}\BibitemShut
  {NoStop}%
\bibitem [{\citenamefont {Kohn}\ \emph {et~al.}(2020)\citenamefont {Kohn},
  \citenamefont {Silvi}, \citenamefont {Gerster}, \citenamefont {Keck},
  \citenamefont {Fazio}, \citenamefont {Santoro},\ and\ \citenamefont
  {Montangero}}]{kohn2020superfluid}%
  \BibitemOpen
  \bibfield  {author} {\bibinfo {author} {\bibfnamefont {L.}~\bibnamefont
  {Kohn}}, \bibinfo {author} {\bibfnamefont {P.}~\bibnamefont {Silvi}},
  \bibinfo {author} {\bibfnamefont {M.}~\bibnamefont {Gerster}}, \bibinfo
  {author} {\bibfnamefont {M.}~\bibnamefont {Keck}}, \bibinfo {author}
  {\bibfnamefont {R.}~\bibnamefont {Fazio}}, \bibinfo {author} {\bibfnamefont
  {G.~E.}\ \bibnamefont {Santoro}}, \ and\ \bibinfo {author} {\bibfnamefont
  {S.}~\bibnamefont {Montangero}},\ }\href {\doibase
  10.1103/PhysRevA.101.023617} {\bibfield  {journal} {\bibinfo  {journal}
  {Phys. Rev. A}\ }\textbf {\bibinfo {volume} {101}},\ \bibinfo {pages}
  {023617} (\bibinfo {year} {2020})}\BibitemShut {NoStop}%
\bibitem [{\citenamefont {Kloss}\ \emph {et~al.}(2020)\citenamefont {Kloss},
  \citenamefont {Reichman},\ and\ \citenamefont {Bar~Lev}}]{kloss2020studying}%
  \BibitemOpen
  \bibfield  {author} {\bibinfo {author} {\bibfnamefont {B.}~\bibnamefont
  {Kloss}}, \bibinfo {author} {\bibfnamefont {D.}~\bibnamefont {Reichman}}, \
  and\ \bibinfo {author} {\bibfnamefont {Y.}~\bibnamefont {Bar~Lev}},\ }\href
  {\doibase 10.21468/SciPostPhys.9.5.070} {\bibfield  {journal} {\bibinfo
  {journal} {SciPost Phys.}\ }\textbf {\bibinfo {volume} {9}},\ \bibinfo
  {pages} {070} (\bibinfo {year} {2020})}\BibitemShut {NoStop}%
\bibitem [{\citenamefont {Strand}\ \emph {et~al.}(2021)\citenamefont {Strand},
  \citenamefont {Vroylandt},\ and\ \citenamefont
  {Gingrich}}]{strand2021companion}%
  \BibitemOpen
  \bibfield  {author} {\bibinfo {author} {\bibfnamefont {N.~E.}\ \bibnamefont
  {Strand}}, \bibinfo {author} {\bibfnamefont {H.}~\bibnamefont {Vroylandt}}, \
  and\ \bibinfo {author} {\bibfnamefont {T.~R.}\ \bibnamefont {Gingrich}},\
  }\href@noop {} {\bibfield  {journal} {\bibinfo  {journal} {arXiv:2201.04107}\
  } (\bibinfo {year} {2021})},\ \Eprint {http://arxiv.org/abs/2201.04107}
  {2201.04107} \BibitemShut {NoStop}%
\bibitem [{\citenamefont {Pippan}\ \emph {et~al.}(2010)\citenamefont {Pippan},
  \citenamefont {White},\ and\ \citenamefont {Evertz}}]{pippan2010efficient}%
  \BibitemOpen
  \bibfield  {author} {\bibinfo {author} {\bibfnamefont {P.}~\bibnamefont
  {Pippan}}, \bibinfo {author} {\bibfnamefont {S.~R.}\ \bibnamefont {White}}, \
  and\ \bibinfo {author} {\bibfnamefont {H.~G.}\ \bibnamefont {Evertz}},\
  }\href {\doibase 10.1103/PhysRevB.81.081103} {\bibfield  {journal} {\bibinfo
  {journal} {Phys. Rev. B}\ }\textbf {\bibinfo {volume} {81}},\ \bibinfo
  {pages} {081108} (\bibinfo {year} {2010})}\BibitemShut {NoStop}%
\bibitem [{\citenamefont {Hubig}\ \emph {et~al.}(2015)\citenamefont {Hubig},
  \citenamefont {McCulloch}, \citenamefont {Schollw{\"o}ck},\ and\
  \citenamefont {Wolf}}]{hubig2015strictly}%
  \BibitemOpen
  \bibfield  {author} {\bibinfo {author} {\bibfnamefont {C.}~\bibnamefont
  {Hubig}}, \bibinfo {author} {\bibfnamefont {I.~P.}\ \bibnamefont
  {McCulloch}}, \bibinfo {author} {\bibfnamefont {U.}~\bibnamefont
  {Schollw{\"o}ck}}, \ and\ \bibinfo {author} {\bibfnamefont {F.~A.}\
  \bibnamefont {Wolf}},\ }\href {\doibase 10.1103/PhysRevB.91.155115}
  {\bibfield  {journal} {\bibinfo  {journal} {Phys. Rev. B}\ }\textbf {\bibinfo
  {volume} {91}},\ \bibinfo {pages} {155115} (\bibinfo {year}
  {2015})}\BibitemShut {NoStop}%
\bibitem [{\citenamefont {Fishman}\ \emph {et~al.}(2020)\citenamefont
  {Fishman}, \citenamefont {White},\ and\ \citenamefont
  {Stoudenmire}}]{itensor}%
  \BibitemOpen
  \bibfield  {author} {\bibinfo {author} {\bibfnamefont {M.}~\bibnamefont
  {Fishman}}, \bibinfo {author} {\bibfnamefont {S.~R.}\ \bibnamefont {White}},
  \ and\ \bibinfo {author} {\bibfnamefont {E.~M.}\ \bibnamefont
  {Stoudenmire}},\ }\href@noop {} {\bibfield  {journal} {\bibinfo  {journal}
  {arXiv:2007.14822}\ } (\bibinfo {year} {2020})},\ \Eprint
  {http://arxiv.org/abs/2007.14822} {arXiv:2007.14822} \BibitemShut {NoStop}%
\bibitem [{\citenamefont {Anderson}(2007)}]{anderson2007modified}%
  \BibitemOpen
  \bibfield  {author} {\bibinfo {author} {\bibfnamefont {D.~F.}\ \bibnamefont
  {Anderson}},\ }\href {\doibase 10.1063/1.2799998} {\bibfield  {journal}
  {\bibinfo  {journal} {J. Chem. Phys.}\ }\textbf {\bibinfo {volume} {127}},\
  \bibinfo {pages} {214107} (\bibinfo {year} {2007})}\BibitemShut {NoStop}%
\end{thebibliography}%

\end{document}